\documentclass[12pt]{article}
\usepackage{amssymb}
\usepackage{graphicx}
\usepackage{graphics}
\usepackage{cite}
\usepackage{ulem}
\usepackage{amsmath}
\usepackage{mathtools}

\def\l{\left(}                                    
\def\r{\right)}

\newcommand{\be}{\begin{equation}}
\newcommand{\ee}{\end{equation}}
\newcommand{\ba}{\begin{align}}
\newcommand{\ea}{\end{align}}
\newcommand{\bg}{\begin{gather}}
\newcommand{\eg}{\end{gather}}
\newcommand{\bseq}{\begin{subequations}}
\newcommand{\eseq}{\end{subequations}}

\begin{document}
\title{Polarization of photons emitted by decaying dark matter}

\author{W. Bonivento$^{1}$, D. Gorbunov$^{2,3}$,     M. Shaposhnikov$^{4}$     and  A. Tokareva$^{2,4}$\\
\mbox{}$^{1}${\small Sezione INFN di Cagliari, Cagliari, Italy}\\ 
\mbox{}$^{2}${\small Institute for Nuclear Research of Russian Academy of
  Sciences, 117312 Moscow,
  Russia}\\  
\mbox{}$^{3}${\small Moscow Institute of Physics and Technology, 
141700 Dolgoprudny, Russia}\\ 
 \mbox{}$^{4}${\small Ecole Polytechnique Fб╢edб╢erale de Lausanne, CH-1015, Lausanne, Switzerland}\\ 
}

\maketitle

\begin{abstract} 
Radiatively decaying dark matter may be searched through investigating 
the photon spectrum of galaxies and galaxy clusters. We
explore whether the properties of dark matter can be constrained
through the study of a polarization state of emitted photons. Starting from the
basic principles of quantum mechanics we show that the models of
symmetric dark matter are indiscernible by the photon
polarization. However, we find that the asymmetric dark matter
consisted of Dirac fermions is a source of circularly polarized
photons, calling for the experimental determination of the photon state.

\end{abstract}

\section{Introduction}

Dark matter (DM) particles can be unstable with a life-time exceeding that of
 the Universe. If these particles have a radiative decay mode, the emission
 of an almost monochromatic photon is a specific signature allowing to 
search for them in astrophysical observations. Similar imprint can also
 come from two-photon annihilation of DM states. The examples of DM 
particles producing  photon(s)  include  sterile neutrino with the mass in the 
keV range \cite{DW,resonant,dolgov,Abazajian1,Abazajian2} (see also 
\cite{Boyarsky:2009ix,Adhikari:2016bei} for reviews), axions and axionlike 
particles \cite{axion1,Jaeckel:2014qea,axion2}, sgoldstinos \cite{Demidov:2014hka},  majorons \cite{majoron}, axinos \cite{axino}, gravitinos \cite{Bomark:2014yja}, self-annihilating particles \cite{annihilation} and 
transitions between two dark matter states \cite{excited}. Some  
(controversial)  indications in favour of these types of DM particles came recently from the analysis of photon spectra emitted by different 
astrophysical objects in X-ray region (an unidentified 3.5 keV line) and were reported in  \cite{Bulbul:2014sua, Boyarsky:2014jta}\footnote{For the current status of this line and future prospects see \cite{Boyarsky:2012rt, Neronov:2016wdd,Adhikari:2016bei}. }.

This unidentified X-ray line may have no connection with the
dark sector  and come from some atomic transition, see the
discussion \cite{Jeltema:2014qfa}. Namely, there are several atomic lines near 
the 3.5 keV feature: potassium K XVIII lines at 3.48 and 3.52 keV, and charge exchange induced line 
of sulfur, S XVI at 3.47 keV. The situation remains unclear because it is hard to estimate 
accurately the flux of photons coming from these transitions, given the lack of precise knowledge 
of the chemical composition and temperature of the cosmic plasma. However, future 
observations with enhanced spectral resolution should be able to distinguish dark matter and atomic lines.

But one can address a question: even if the dark matter nature of the line would be proved, how can we decide what type of dark matter particle produces this line? For example, can we
distinguish between a boson and a fermion? Of course, if we had a
possibility to catch and study all decay products of hypothetical dark
matter particle then we would be able to determine the spin and parity
of the original particle. Unfortunately, we can register only one
particle---photon---because another decay product flies in the
opposite direction.

Clearly, besides the energy photons may carry another information
encoded in their polarization state. The aim of this paper is to
address the question whether the polarization measurements can help to
constrain the properties of the dark matter\footnote{For discussion of photon polarizations in collider experiments and in indirect searches see \cite{Fichet:2016clq} and \cite{Garcia-Cely:2016pse,Ibarra:2016fco}, respectively.}. We will show that the quantum
mechanical state of each arriving individual photon may be different
for the different types of dark matter particles. Still, we will
demonstrate that it is in principle not
possible to make such measurements that could determine the spin of
dark matter particle if it is symmetric (i.e. contains equal numbers
of particles and antiparticles).

We will show, however, that the case of asymmetric dark matter
consisting of Dirac fermions is different. Namely, it provides a
circularly polarized photons when decaying to the photon 
and the fermion (e.g. neutrino). This
circular polarization, in principle, could be detected though it may
be a challenge for the real experiment. We will discuss shortly a few
models which can lead to radiatively decaying asymmetric dark matter.

The paper is organised as follows. In Section 2, we obtain the single
photon polarization state for decaying scalar, fermion (with spin 1/2
and 3/2) and for the atomic line and prove that it is impossible to
distinguish between them if dark matter is symmetric. In Section 3 we
show that in a case of asymmetric dark matter consisted of Dirac
fermions one obtains the circularly polarized emission line. In
Section 4 more complicated models including the transition between
dark states are briefly discussed and the common criteria for the
polarized line are formulated. The last section is conclusions.

\section{Symmetric dark matter}
\label{symmetric}

In this section we determine the quantum state of the photon emitted in dark matter decay. We assume that DM is CP-symmetric and that it can disintegrate directly (i.e.  without any intermediate states) into final states with photon(s).   In order to analyse different dark matter models we use the language of the effective field theory. We order our account by the spin of DM particle: starting from the scalar we go to the spin 3/2 fermion and complete with a discussion of atomic transitions. Then we consider possibilities to distinguish between different spins of dark matter particle.

\subsection{Scalar and pseudoscalar}
The neutral scalar $\phi$ and pseudoscalar 
$a$ particles may interact with the
photon through the lowest order operators 
\be L=\frac{1}{\Lambda}\phi
F_{\mu \nu}F^{\mu \nu},\quad L=\frac{1}{\Lambda} a F_{\mu
  \nu}\tilde{F}^{\mu \nu}, 
\ee 
respectively. Here $\tilde{F}_{\mu
  \nu}=\epsilon_{\mu \nu\alpha\beta}F^{\alpha\beta}/2$ is the dual
electromagnetic field tensor and $\Lambda$ is a mass parameter.  In
order to obtain the reasonable decay width providing the observable X-ray line
intensity one needs $\Lambda$ to be of order Planck scale
\cite{Krall:2014dba} which looks natural for the variety of axion and
axionlike particle models \cite{axion1,Jaeckel:2014qea,axion2}.

The spin-0 boson decays to two photons which polarization state is the maximally entangled Bell state \cite{Sudbery:1986sr}:
\be 
\label{Bell}
|\Psi\rangle = \frac{1}{\sqrt{2}}\l |LL\rangle \pm |RR\rangle \r.
\ee
Hereafter we use the notation $|L\rangle$ for the left-handed photon
(with the spin projection -1 on the momentum direction) and
$|R\rangle$ for the right-handed photon. $|RR\rangle$ means, for
example, the state of two right-handed photons. The sign in
\eqref{Bell} is determined by the parity of initial particle: plus for
the scalar and minus for the pseudoscalar.

We can register only one photon from each pair because the other one
flies in the opposite direction. In order to obtain the reduced
density matrix for one of the photons, we need to take a trace of
 the density matrix $|\Psi\rangle\langle\Psi|$ 
over all states of the unobservable photon. The result is
the unity density matrix corresponding to the maximally mixed and
unpolarized state for every coming photon in the flux
\cite{Sudbery:1986sr}: 
\be 
\label{scalar}
\rho=\frac{1}{2}(|R\rangle\langle R|+|L\rangle\langle L|).
\ee

\subsection{Fermion with spin $1/2$}

An extra singlet Majorana fermion $N$ (sterile neutrino) may be added to the Standard
Model (SM) in order to explain dark matter. The only
renormalizable interaction between $N$ and SM particles allowed by
symmetries may be written as 
\be 
\label{4}
L= f\,\bar{l_L}N
\,\tilde{{\cal H}}+ h. c.~.  
\ee 
Here $\tilde{{\cal H}}=\epsilon {\cal H^*}$, where $\epsilon$ is
$2\times2$ antisymmetric unit matrix and ${\cal H}$ is the Higgs
  doublet, 
$l_L$ the left SM lepton doublet, $f$ is a small dimensionless
coupling. This interaction at one loop level leads to the effective
coupling between the sterile neutrino and the field strength $B_{\mu\nu}$ of 
the $U(1)_Y$ gauge boson: 
\be L\;\propto\;\frac{\tilde{{\cal H}}}{\Lambda^2}
\,\bar{l}_L\sigma_{\mu\nu} N \,B^{\mu\nu} + h.c.~.  
\ee 
Here
$\sigma_{\mu\nu}=(\gamma_{\mu}\gamma_{\nu}-\gamma_{\nu}\gamma_{\mu})/2$. The
scale $\Lambda$ is connected with the parameter $f$ in \eqref{4} and Higgs
vacuum expectation value $v$ as $\Lambda\sim v/\sqrt{f}$. After 
spontaneous symmetry breaking one obtains the operator describing the
interaction between the extra fermion $N$, SM neutrino $\nu_L$ and
photon $F_{\mu\nu}$: 
\be 
L\;\propto\; \frac{v}{\Lambda^2}
\,\bar{\nu}_L\sigma_{\mu\nu} N \,F^{\mu\nu} + h.c.  
\ee

Since the SM (active) neutrinos are 
always in a pure spin state (left-handed for
neutrinos and right-handed for antineutrinos) we expect that the
photon state is determined by the initial state of sterile
neutrino. Clearly, it depends on the mechanism of sterile neutrino
production in the early Universe. If they were produced 
in the process of the scalar (inflaton, majoron) decay
\cite{Shaposhnikov:2006xi,Kusenko:2006rh} then each particle is
expected to be in the maximally mixed state and the same holds for the
photon. So we obtain the density matrix \eqref{scalar}. However, if
the Majorana particles were produced by oscillations in lepton symmetric 
\cite{DW} or lepton asymmetric plasma \cite{resonant} then we
expect that each particle is in the pure spin state but the spin
vectors of different particles are distributed randomly. Each particle
may decay to the SM neutrino and antineutrino with equal
probability (neglecting the possible $CP$-violation in \eqref{4}). 
Then, the state of photon $|\gamma\rangle$ may be
described as follows: 
\be
\label{fermion_M}
\begin{matrix}
|\gamma\rangle=\left\{ 
\begin{matrix} \sqrt{1-\beta}|L\rangle +\sqrt{\beta}e^{i\alpha}|R\rangle,~\text{with probability } \frac{1}{2},\\ \sqrt{1-\beta}|R\rangle +\sqrt{\beta}e^{i\alpha}|L\rangle,~\text{with probability } \frac{1}{2} .
\end{matrix} \right.
\end{matrix}
\ee
Here the parameter $\beta$ depends on the direction of the Majorana
particle's spin and, therefore, is a random parameter. The same is for
the phase $\alpha$ since we expect no spin correlations for the dark
matter particles. We see that the polarization state of each
individual photon is a pure state in the quantum mechanical sense.

If the  DM fermion $\psi$ is of the Dirac type (we use this
notation for the Dirac particle instead of $N$ left for the Majorana case)
then the particle decays only to the left-handed
neutrino and antiparticle decays 
to the right-handed antineutrino, correspondingly. If $s$ stands for the
momentum projection of the photon spin ($s=1$ for the right-handed photon and
$s=-1$ for the left-handed one) 
then we find the fermion and antifermion
decay width to be
\be 
\label{gamma}
\Gamma_{\psi}=\Gamma_0(1-s),\quad \Gamma_{\bar{\psi}}=\Gamma_0(1+s),
\quad \Gamma_0  \;\propto\;\frac{v^2 m_{\psi}^3}{\Lambda^4}.
\ee
We see that the fermion provides only right-handed photons while the
antifermion gives only left-handed ones. 
Every photon is in the pure polarization
state. If the number of fermions equals the number of
  antifermions,  
the photon flux will consist of equal numbers of left-handed and right-handed
states:
\be 
\label{fermion}
\begin{matrix}
|\gamma\rangle=\left\{ 
\begin{matrix} |R\rangle,~\text{with probability } \frac{1}{2},\\ |L\rangle,~\text{with probability } \frac{1}{2} .
\end{matrix} \right.
\end{matrix}
\ee

\subsection{Fermion with spin $3/2$}

We omit the discussion on the vector dark matter particles with spin 1
because the vector particle can not decay to two photons. Therefore we
go further and consider spin 3/2 fermion. This particle of Majorana
type naturally arises in supergravity models as a graviton
superpartner -- gravitino. The
effective interaction of spin 3/2 fermion $\psi_{\rho}$ with photon
may be written as \be L=\frac{v}{\Lambda^2}
\,\bar{\psi_{\rho}}\sigma_{\mu\nu}\gamma^{\rho} \nu_L \,F^{\mu\nu} +
h. c.~.  \ee For the Majorana particle, all statements of the previous
section related to the spin $1/2$ remain unchanged. Namely, if the
gravitino was produced in the mixed state then we register each photon
in the state \eqref{scalar} while for the pure states we have
\eqref{fermion_M}.

If the spin $3/2$ particle is Dirac (the corresponding model is worked
out in \cite{Dutta:2015ega}) then we again obtain results which are
independent of the production mechanism. Namely, the spin-averaged
decay width depends on the photon polarization as
\be 
\label{gamma_3/2}
\Gamma_{\psi_{\mu}}=\Gamma_{3/2}(1+s),\quad \Gamma_{\bar{\psi}_{\mu}}=\Gamma_{3/2}(1-s), \quad \Gamma_{3/2}\;\propto\;\frac{v^2 m_{\psi}^3}{\Lambda^4}.
\ee
We see that the particle provides only right-handed photons while
antiparticle gives only left-handed ones. 
Then the photon state in the symmetric
case is the same as for spin $1/2$ fermions \eqref{fermion}.

\subsection{Atomic transition}

 The 3.5 keV line received many interpretations in terms of
dark matter particle decay, but it may well be a result 
of usual atomic transitions \cite{Jeltema:2014qfa}. 
Though this case may be distinguished with the enhanced spectral
resolution we also discuss a polarization state of photons in the case
of atomic transitions.

In a case of the line corresponding to the dipole atomic transition
(i. e. the transition when the orbital quantum number of electron
changes by unity) the polarization of emitted photon depends on the
change of magnetic quantum number which may be $\Delta m=-1,~1,~
0$. These cases correspond to the emission of left, right and linearly
polarized photon, correspondingly. The textbook knowledge (see, for
example, \cite{sobelman}) predicts that all three cases $\Delta
m=0,~\pm 1$ happen with equal probability if initial atoms are
unpolarized (this condition holds in the interstellar
medium). Therefore, the state of photon may be described as 
\be
\label{atom}
\begin{matrix}
|\gamma \rangle=\left\{ 
\begin{matrix*}[l] |R\rangle,~\text{with probability } \frac{1}{3},\\
  |L\rangle,~\text{with probability } \frac{1}{3}, \\
  \frac{1}{\sqrt{2}}(|R\rangle+e^{i \alpha} |L\rangle),~ \text{with probability } \frac{1}{3}.
\end{matrix*} \right.
\end{matrix}
\ee
Here $\Psi=(|R\rangle+e^{i \alpha} |L\rangle)/2$ is the linearly
polarized state and the phase $\alpha$ is a random number
characterising the direction of linear polarization. We see that the
state of photons coming from the dipole atomic transition differs from all cases
listed above\footnote{Multipole transitions are expected to provide more complicated state distribution but, as we will show in Section 2.5, only the density matrix averaged over all the photon states really makes sense. Thus, the final result would be the same as in dipole transitions.}. In the next section we discuss whether this difference
is measurable in any possible experiment.

\subsection{How to distinguish?}

In the real experiment, we register a flux of photons. It can be
described by the average polarization density matrix (see, for
example, \cite{blum} for the definition) even when each individual
photon is in the pure state. One can see that for all cases described in the previous sections the averaged density matrix for the flux of photons is
the same and takes the form \eqref{scalar}. But we deal with different
states of individual photons: in the case of fermion, we have the pure
state while for the scalar we have the completely mixed state. For the
atomic transition, some photons have the linear polarization. If we were dealing with a {\it known} pure state,  one can easily discern it from the mixed state. But in the  case  under consideration the structure of the pure state is {\it unknown} as it depends on the spin projection of the DM particle. Is it
possible to distinguish between the described cases in any type of experiment?

{\it Single photon measurements.} The only thing that we can measure
for a single photon is its projection on the basic state. After the
first measurement, the state collapses to its projection and the
initial state is lost. In order to reconstruct the full density matrix
one need to make at least three projection measurements \cite{Kwiat}
which is impossible to do for a single photon.

{\it Multiple photon measurements.} Potentially, we can imagine to
collect many photons from the coming flux and hold them in a
box. Then, any possible measurement may be reduced to finding the
correlation function for the corresponding product of operators:
$A=O(x_1)O(x_2)\dots O(x_n)$. Here $O$ stands for any gauge invariant
operator containing photon field and $x_1\dots x_n$ are the different
points of the space and time. These correlators would be different for the
pure and mixed many-particle states but each pure state correlator has
its own dispersion defined by the usual formula 
\be
D_A(\Psi)=\langle\Psi| (A-\langle A\rangle)^2
|\Psi\rangle=\langle\Psi| A^2 |\Psi\rangle-(\langle A\rangle)^2, 
\ee
where $\langle A\rangle=\langle\Psi| A |\Psi\rangle$. The dispersion
of the correlator for the mixed state with density matrix $\rho$ is
\be 
\label{14}
D_A(\rho)=\text{Tr}(\rho A^2)-(\text{Tr}(\rho A))^2.  
\ee 
Then, the theorem proven in \cite{Lloyd} provides the answer to the question
about the physical difference between the pure and mixed state for our case when density matrices averaged over all photons are the same.\\ 
{\it  Theorem.}\\ 
Let $A$ be an hermitian operator corresponding to the
measurement of some quantity and $N$  be the number of particles
(i.e. the number of photons), $|\Psi\rangle$ be a pure state,
$\rho=2^{-N} \times {\bf 1}$  be the density matrix describing the
completely mixed state \footnote{This means that each collected photon
  is equally likely left-handed and right-handed.}. Then the
difference between the correlators over the pure state and over the state
described by the density matrix $\rho$ is always smaller (by factor
$1/\sqrt{N+1}$) than the intrinsic dispersion of this correlator \eqref{14},
calculated over the density matrix $\rho$: 
 \be
\l\langle\Psi|A|\Psi\rangle_{|\Psi\rangle} - \text{Tr}(\rho
A)\r^2=\frac{D_A(\rho)}{N+1}.  
\ee 

The left-hand
side means the averaged over all possible pure states $|\Psi\rangle$
squared difference between the correlators for the pure and mixed
state. In other words, this theorem reflects the fact that each correlator for the mixed state has a relatively large variance. So, one can not decide whether the measured correlation function corresponds to the pure or mixed state because the results for the pure states lie in the band of possible values for the mixed state. This implies that one can never distinguish between the indefinite pure state and mixed state of photons.

Summarising, we have no chance to determine if the photon coming from
dark matter decay is in the pure state. So we can't distinguish between
cases leading to the equal density matrices when averaging over all
photon flux. By this reason, the polarization would not help us to
detect even the case of the line provided by the usual atomic
transition: the distribution \eqref{atom} leads again to the unit density matrix when averaged over the photon flux.

\section{Asymmetric dark matter}

In this section, we show that the study of the dark matter photons
polarization is still important because it allows detecting the
asymmetry between the number of fermions and anti-fermions constituting the dark matter if it has a Dirac fermionic nature.  As an example, we
consider models connected with the sterile neutrino dark matter
\cite{DW,resonant, dolgov, Abazajian1, Abazajian2,
  Boyarsky:2009ix}. Usually, sterile neutrinos are treated as Majorana
particles but it is also possible to consider them to be Dirac
fermions. Then, the asymmetry in the dark sector strongly depends on
the production mechanism in the early Universe.  In case of
  thermal production in plasma no significant asymmetry is expected.

However, the asymmetry may arise in the processes sharing some
features of leptogenesis.  An example is provided by the resonant 
production of sterile neutrinos \cite{resonant} originally proposed
for the Majorana neutrinos. The latter process
would be still valid for the Dirac case
as well. This scenario works as follows. Due to the presence of the SM
lepton asymmetry the dispersion relation of the sterile neutrino is
modified in such a way that at some temperature the level crossing
with SM neutrinos happens. At this moment the large part of SM
neutrinos presented in the Universe converts into sterile ones. Since
in the Dirac case fermions are mixed only with fermions,  but not
with antifermions, one  obtains that the major part of
lepton asymmetry in the SM neutrino
sector directly converts to the asymmetry in the dark sector. The
described mechanism works only for large lepton asymmetry
$\eta_L=(n_L-\bar{n}_L)/(n_L+\bar{n}_L)>10^{-4}$ and it can naturally
provide $\eta_{\psi}=(n_{\psi}-\bar{n}_{\psi})/(n_{\psi}+\bar{n}_{\psi})\sim 1$ (where $n_{\psi}$ and $\bar{n}_{\psi}$ are the density numbers of DM particles and antiparticles, correspondingly), depending on the choice of parameters.

Since the particle decay may provide only  left-handed
photons while antiparticle give only right-handed photons (see
eq. \eqref{gamma}) the asymmetric case leads to the polarization
density matrix of the flux in the basis $(|R\rangle,~|L\rangle)$ of
the form: 
\be
\label{1/2}
\rho=\frac{1}{n+\bar{n}}\l\begin{matrix}
\bar{n} & 0 \\
0 & n
\end{matrix}
\r=\frac{1}{2}\l 1-\eta_{\psi} \sigma_3 \r,
\ee
where $\sigma_3=|R\rangle\langle R|-|L\rangle\langle L|$ is the third
Pauli matrix. If $\eta_{\psi}\ne 1$ the flux is partially polarized
corresponding to the set of Stokes parameters $S_1=S_2=0,~
S_3=-\eta_{\psi}$ (for parametrisation of polarized light see, for
example, \cite{Kwiat}).

To complete the consideration let us study the case of the Dirac
fermion with spin $3/2$. To the best of our knowledge, no dark matter
models assuming asymmetry of spin-$3/2$ fermions were suggested in the
literature but this does not look to be impossible. In 
Section\,\ref{symmetric}  we
found that decaying fermions of spin $3/2$ yield only 
  right-handed photons
while antifermions yield only left-handed ones. 
Therefore, if there is an
asymmetry $\eta_{\psi}$ one can detect the polarization density matrix
for coming flux of photons to be
\be 
\label{3/2}
\rho=\frac{1}{2}\l 1+\eta_{\psi} \sigma_3 \r.
\ee
The difference in signs in \eqref{1/2} and \eqref{3/2} is due to the
fact that particles with spin 3/2 decay to the right-handed photons 
while  spin-1/2 particles yield the left-handed ones.

While it is clear how to detect the circular polarization of the
visible light, for many other wave-ranges, in particular for 
the X-ray bandwidth, it is not so obvious. The
Compton scattering, which detects the linear polarization of X-rays,
does not distinguish between right and left circular
polarizations. However, there are some attempts to measure such kind
of X-ray polarization in laboratory \cite{kabachnik} that use circular dichroism in two photon ionisation of helium. To the best of our knowledge, no methods of detecting circular polarization of gamma-rays have been suggested yet. However, in this bandwidth there are more possibilities for indirect searches due to the appearance of other decay channels (see for example \cite{Garcia-Cely:2016pse,Ibarra:2016fco}). In any case,
detection of the circular polarization of the dark matter photons in
future experiments may be a very clear signature of the fermionic
nature and asymmetry in the dark matter sector.
     
\section{Dark state transition models}

Besides the DM particle decay, the line-like feature may also be produced in transitions between two dark states \cite{excited}. In models of
such type, the dark matter goes to the excited state due to the plasma
collisions or background emission and then decays back emitting the photon. If the transition happens between the two Dirac fermions
$\psi$ and $\chi$ the photons would be polarized only when the following
conditions are satisfied:
\begin{enumerate}
\item The number density of heavier state, $\psi$, differs from the number of $\bar{\psi}$.
\item The interaction with the lighter state $\chi$ is $P$-asymmetric.
\end{enumerate}
Obviously, for decays of $\psi$ to SM neutrino and photon the second
condition is satisfied automatically providing the results for the
photons polarization obtained in the previous sections.

If transition happens between the two bosonic states (scalar and pseudoscalar or vector
\cite{Farzan:2014foo}) then, clearly, no circular polarization is
expected because the corresponding process is  $P$- and $C$-
symmetric. Although, as far as we know, the models satisfying both
conditions listed above are not discussed in the literature it looks
not impossible to construct them. But for the most cases yet
considered the emission would be unpolarized.

\section{Conclusions}

Unfortunately, the most important symmetric dark matter models exhibiting the photon line
-- fermion (sterile neutrino) and
scalar (axion, ALP) -- are indistinguishable in observations of the
line polarization. So, in order to decide what kind of dark matter
particle produces the X-ray line supplementary experiments will be
needed. For example, sterile neutrino dark matter might be searched in
the tritium decay due to its small mixing with the SM neutrinos
\cite{Bezrukov:2006cy}. The axions and axionlike particles may be
probed in a variety of special laboratory experiments connected with
the axion-photon oscillations in the magnetic field, see
\cite{Jaeckel:2014qea} for a review and references.

However, if the experiment will show the circularly polarized line it
would be a smoking gun of its dark matter origin because there is no
way to obtain such polarization in usual atomic transitions. Moreover,
it would be a signature of asymmetric fermionic nature of dark matter
and the amount of polarization would be directly connected with the
asymmetry in the dark sector. By this reason, we underline the
importance of polarisation studies in the future observations.

 In this paper we concentrated mostly on the keV-scale photons,
motivated by the recent indications of 3.5\,keV line. However, all the
considerations are valid for any photon energy range.

\vskip 0.2cm
This work was supported by the ERC-AdG-2015 grant 694896.
We thank A.\,Grum-Grzhimailo, V.\,Savona and O.\,Tikhonova for 
correspondence, discussions and consultations in the
field of quantum optics and atomic physics, and A. \,Boyarsky and O. \,Ruchayskiy for helpful comments. The work of AT and DG
was supported by Russian Science Foundation grant
14-12-01430 in the part where quantum mechanical properties of the
produced photons are studied. The work of MS was
supported partially by the Swiss National Science Foundation. 



\begin{thebibliography}{99}

\bibitem{DW} 
  S.~Dodelson and L.~M.~Widrow,
  Phys.\ Rev.\ Lett.\  {\bf 72}, 17 (1994)
  [hep-ph/9303287].
    \bibitem{resonant}
  X.~D.~Shi and G.~M.~Fuller,
  Phys.\ Rev.\ Lett.\  {\bf 82}, 2832 (1999)
  [astro-ph/9810076].
 \bibitem{dolgov} 
A.~D.~Dolgov and S.~H.~Hansen,
  Astropart.\ Phys.\  {\bf 16}, 339 (2002)
  [hep-ph/0009083].
    \bibitem{Abazajian1} 
  K.~Abazajian, G.~M.~Fuller and M.~Patel,
  Phys.\ Rev.\ D {\bf 64}, 023501 (2001)
  [astro-ph/0101524].
  \bibitem{Abazajian2} 
  K.~Abazajian, G.~M.~Fuller and W.~H.~Tucker,
  Astrophys.\ J.\  {\bf 562}, 593 (2001)
  [astro-ph/0106002].
  
\bibitem{Boyarsky:2009ix} 
  A.~Boyarsky, O.~Ruchayskiy and M.~Shaposhnikov,
  Ann.\ Rev.\ Nucl.\ Part.\ Sci.\  {\bf 59}, 191 (2009)
  [arXiv:0901.0011 [hep-ph]].
\bibitem{Adhikari:2016bei} 
R.  Adhikari  {\it et al.},
  [arXiv:1602.04816 [hep-ph]].
  

  
\bibitem{axion1}
 T.~Higaki, K.~S.~Jeong and F.~Takahashi,
  Phys.\ Lett.\ B {\bf 733}, 25 (2014)
  [arXiv:1402.6965 [hep-ph]].
  \bibitem{Jaeckel:2014qea} 
  J.~Jaeckel, J.~Redondo and A.~Ringwald,
  Phys.\ Rev.\ D {\bf 89}, 103511 (2014)
   [arXiv:1402.7335 [hep-ph]].
  \bibitem{axion2}
       H.~M.~Lee, S.~C.~Park and W.~I.~Park,
  Eur.\ Phys.\ J.\ C {\bf 74}, 3062 (2014)
  [arXiv:1403.0865 [astro-ph.CO]]. 
  

\bibitem{Demidov:2014hka}
  S.~V.~Demidov and D.~S.~Gorbunov,
  Phys.\ Rev.\ D {\bf 90} (2014) 035014
  [arXiv:1404.1339 [hep-ph]].

\bibitem{majoron} 
  F.~S.~Queiroz and K.~Sinha,
  Phys.\ Lett.\ B {\bf 735}, 69 (2014)
  [arXiv:1404.1400 [hep-ph]].  


\bibitem{axino}  
J.~C.~Park, S.~C.~Park and K.~Kong,
  Phys.\ Lett.\ B {\bf 733}, 217 (2014)
  [arXiv:1403.1536 [hep-ph]].

\bibitem{Bomark:2014yja} 
  N.-E.~Bomark and L.~Roszkowski,
  Phys.\ Rev.\ D {\bf 90}, 011701 (2014)
  [arXiv:1403.6503 [hep-ph]].

  
\bibitem{annihilation}
  E.~Dudas, L.~Heurtier and Y.~Mambrini,
  Phys.\ Rev.\ D {\bf 90}, 035002 (2014)
  [arXiv:1404.1927 [hep-ph]].
\bibitem{excited}  
  D.~P.~Finkbeiner and N.~Weiner,
  arXiv:1402.6671 [hep-ph].  
  J.~M.~Cline, Y.~Farzan, Z.~Liu, G.~D.~Moore and W.~Xue,
  Phys.\ Rev.\ D {\bf 89} (2014) 121302
  [arXiv:1404.3729 [hep-ph]].
    H.~Okada and T.~Toma,
  Phys.\ Lett.\ B {\bf 737}, 162 (2014)
  [arXiv:1404.4795 [hep-ph]].
  C.~Q.~Geng, D.~Huang and L.~H.~Tsai,
  JHEP {\bf 1408}, 086 (2014)
  [arXiv:1406.6481 [hep-ph]].  F.~D'Eramo, K.~Hambleton, S.~Profumo and T.~Stefaniak,
  Phys.\ Rev.\ D {\bf 93}, no. 10, 103011 (2016)
  [arXiv:1603.04859 [hep-ph]]. J.~P.~Conlon, F.~Day, N.~Jennings, S.~Krippendorf and M.~Rummel,
  arXiv:1608.01684 [astro-ph.HE].
\bibitem{Bulbul:2014sua} 
  E.~Bulbul, M.~Markevitch, A.~Foster, R.~K.~Smith, M.~Loewenstein and S.~W.~Randall,
  Astrophys.\ J.\  {\bf 789}, 13 (2014)
  [arXiv:1402.2301 [astro-ph.CO]].
  
\bibitem{Boyarsky:2014jta} 
  A.~Boyarsky, O.~Ruchayskiy, D.~Iakubovskyi and J.~Franse,
  Phys.\ Rev.\ Lett.\  {\bf 113}, 251301 (2014)
  [arXiv:1402.4119 [astro-ph.CO]].

\bibitem{Boyarsky:2012rt} 
  A.~Boyarsky, D.~Iakubovskyi and O.~Ruchayskiy,
  Phys.\ Dark Univ.\  {\bf 1}, 136 (2012)
  [arXiv:1306.4954 [astro-ph.CO]].
\bibitem{Neronov:2016wdd} 
 A.~Neronov and D.~Malyshev,
  Phys.\ Rev.\ D {\bf 93}, no. 6, 063518 (2016)
  [arXiv:1509.02758 [astro-ph.HE]].
\bibitem{Jeltema:2014qfa} 
  T.~E.~Jeltema and S.~Profumo,
  Mon.\ Not.\ Roy.\ Astron.\ Soc.\  {\bf 450}, no. 2, 2143 (2015)
  [arXiv:1408.1699 [astro-ph.HE]],
   A.~Boyarsky, J.~Franse, D.~Iakubovskyi and O.~Ruchayskiy,
  arXiv:1408.4388 [astro-ph.CO],
  E.~Bulbul, M.~Markevitch, A.~R.~Foster, R.~K.~Smith, M.~Loewenstein and S.~W.~Randall,
  arXiv:1409.4143 [astro-ph.HE],
   T.~Jeltema and S.~Profumo,
  arXiv:1411.1759 [astro-ph.HE],
    T.~E.~Jeltema and S.~Profumo,
  Mon.\ Not.\ Roy.\ Astron.\ Soc.\  {\bf 458}, no. 4, 3592 (2016)
  [arXiv:1512.01239 [astro-ph.HE]],
E.~Carlson, T.~Jeltema and S.~Profumo,
  JCAP {\bf 1502}, no. 02, 009 (2015)
  [arXiv:1411.1758 [astro-ph.HE]],
  D.~Iakubovskyi,
  Mon.\ Not.\ Roy.\ Astron.\ Soc.\  {\bf 453}, no. 4, 4097 (2015)
  [arXiv:1507.02857 [astro-ph.HE]],
C.~Shah, S.~Dobrodey, S.~Bernitt, R.~Steinbrügge, J.~R.~C.~López-Urrutia, L.~Gu and J.~Kaastra,
  arXiv:1608.04751 [astro-ph.HE],
  L.~Gu, J.~Kaastra, A.~J.~J.~Raassen, P.~D.~Mullen, R.~S.~Cumbee, D.~Lyons and P.~C.~Stancil,
  Astron.\ Astrophys.\  {\bf 584}, L11 (2015)
  [arXiv:1511.06557 [astro-ph.HE]],
  S.~Riemer-Sørensen,
  Astron.\ Astrophys.\  {\bf 590}, A71 (2016)
  [arXiv:1405.7943 [astro-ph.CO]],
  K.~J.~H.~Phillips, B.~Sylwester and J.~Sylwester,
  Astrophys.\ J.\  {\bf 809}, 50 (2015),
  V.~K.~Dubrovich,
  Astron.\ Lett.\  {\bf 40}, no. 12, 749 (2014)
  [arXiv:1407.4629 [astro-ph.HE]].
\bibitem{Fichet:2016clq} 
  S.~Fichet,
  arXiv:1609.01762 [hep-ph].
  \bibitem{Garcia-Cely:2016pse} 
  C.~Garcia-Cely and J.~Heeck,
  JCAP {\bf 1608}, no. 08, 023 (2016)
  [arXiv:1605.08049 [hep-ph]].
  
\bibitem{Ibarra:2016fco} 
  A.~Ibarra, S.~Lopez-Gehler, E.~Molinaro and M.~Pato,
  arXiv:1604.01899 [hep-ph].  
  
\bibitem{Krall:2014dba} 
  R.~Krall, M.~Reece and T.~Roxlo,
  JCAP {\bf 1409}, 007 (2014)
  [arXiv:1403.1240 [hep-ph]].
\bibitem{Sudbery:1986sr} 
  A.~Sudbery,
  ``Quantum Mechanics And The Particles Of Nature. An Outline For Mathematicians,''
  Cambridge, Uk: Univ. Pr. (1986).
\bibitem{Shaposhnikov:2006xi} 
  M.~Shaposhnikov and I.~Tkachev,
  Phys.\ Lett.\ B {\bf 639}, 414 (2006)
  [hep-ph/0604236].
\bibitem{Kusenko:2006rh} 
  A.~Kusenko,
  Phys.\ Rev.\ Lett.\  {\bf 97}, 241301 (2006)
  [hep-ph/0609081].
\bibitem{Dutta:2015ega} 
  S.~Dutta, A.~Goyal and S.~Kumar,
  JCAP {\bf 1602}, no. 02, 016 (2016)
  [arXiv:1509.02105 [hep-ph]].
\bibitem{sobelman}
Sobelman, I.I.,
Atomic spectra and radiative transitions,
Springer series in chemical physics (1979)
https://books.google.ru/books?id=5wy1AAAAIAAJ.

\bibitem{blum} 
  K.~Blum,
  ``Density Matrix Theory and Applications,''
  Springer (2012).
\bibitem{Kwiat} 
 D. James, P. Kwiat, W. Munro and A. White,
  Phys.\ Rev.\ A\  {\bf 64}, 052312 (2001).
  
 \bibitem{Lloyd}
S. Lloyd,
Ph.D. Thesis, The Rockfeller University, Chapter 3 (1988)
[arxiv:1307.0378].
  

   \bibitem{kabachnik}
  T. Mazza et al.,
Nature Communications {\bf 5}, 3648 (2014).

  \bibitem{Farzan:2014foo} 
  Y.~Farzan and A.~R.~Akbarieh,
  JCAP {\bf 1411}, no. 11, 015 (2014)
  [arXiv:1408.2950 [hep-ph]].
  
 
\bibitem{Bezrukov:2006cy} 
  F.~L.~Bezrukov and M.~Shaposhnikov,
  Phys.\ Rev.\ D {\bf 75}, 053005 (2007)
  [hep-ph/0611352].


\end{thebibliography}
\end{document}